\documentclass[prd, preprint,nofootinbib,superscriptaddress,12pt]{revtex4-1}
\pdfoutput=1

\newcount\showkey \showkey=0  
\newcount\draft \draft=1 

\usepackage[dvipsnames]{xcolor}
\usepackage{tabularx}

\ifnum\showkey=1
  \usepackage{seqsplit}
  \usepackage[color]{showkeys}
    \definecolor{refkey}{rgb}{0.9, 0.43, 0.63}
    \definecolor{labelkey}{rgb}{0.59, 0.43, 0.63}
  \usepackage{xstring}
  
\fi

\usepackage[pdfusetitle,pdfborder={0 0 0}]{hyperref}
   \hypersetup{colorlinks,
               linkcolor={red!70!black},
               citecolor={green!40!black},
               urlcolor={blue!80!black}}
\usepackage{microtype}   
\usepackage{amsmath}     
  \usepackage{amsthm}    
  \usepackage{amssymb}   
\usepackage{siunitx}     
\usepackage{booktabs}    
\usepackage[capitalize]{cleveref}    
  
  \crefname{section}{Sec.}{Secs.}
  \crefname{appendix}{App.}{Apps.}

\usepackage[shortlabels,inline]{enumitem}  
\usepackage{mathtools} 
\usepackage{subdepth}    

\usepackage{graphicx,multirow,subfigure}
\usepackage{float} 
\usepackage[T1]{fontenc}
\usepackage{physics}
\usepackage{bbold}
\usepackage{rotating} 
\setlist[enumerate,2]{leftmargin=0.45em}
\usepackage{comment}

\DeclareFontFamily{U}{mathx}{\hyphenchar\font45}
\DeclareFontShape{U}{mathx}{m}{n}{<-> mathx10}{}
\DeclareSymbolFont{mathx}{U}{mathx}{m}{n}
\DeclareMathAccent{\widebar}{0}{mathx}{"73}

\newcommand{\beq}{\begin{equation}}
\newcommand{\eeq}{\end{equation}}

\providecommand{\ord}{O}
\providecommand{\ie}{\emph{i.e.}}

\providecommand{\Dbar}{\widebar{D}}
\newcommand{\A}{\mathcal{A}}

\begin{document}

\title{\boldmath Toward extracting \texorpdfstring{$\gamma$}{gamma} from \texorpdfstring{$B\to DK$}{B to DK} without binning}

\author{Jeffrey V. Backus}
\email{jb2134@princeton.edu}
\affiliation{LEPP, Department of Physics, Cornell University, Ithaca, NY 14853, USA}

\author{Marat Freytsis}
\email{mf823@cornell.edu}
\affiliation{LEPP, Department of Physics, Cornell University, Ithaca, NY 14853, USA}
\affiliation{NHETC, Department of Physics and Astronomy, Rutgers University, Piscataway, NJ 08854, USA}

\author{Yuval~Grossman}
\email{yg73@cornell.edu}
\affiliation{LEPP, Department of Physics, Cornell University, Ithaca, NY 14853, USA}

\author{Stefan Schacht}
\email{stefan.schacht@manchester.ac.uk}
\affiliation{Department of Physics and Astronomy, University of Manchester, Manchester M13 9PL, United Kingdom}

\author{Jure Zupan}
\email{zupanje@ucmail.uc.edu}
\affiliation{Department of Physics, University of Cincinnati, Cincinnati, Ohio 45221, USA}

\begin{abstract}
$B^\pm \to DK^\pm$ transitions are known to provide theoretically clean information about the CKM angle $\gamma$, with the most precise available methods exploiting the cascade decay of the neutral $D$ into $CP$ self-conjugate states.
Such analyses currently require binning in the $D$ decay Dalitz plot, while a recently proposed method replaces this binning with the truncation of a Fourier series expansion.
In this paper, we present a proof of principle of a novel alternative to these two methods, in which no approximations at the level of the data representation are required.
In particular, our new strategy makes no assumptions about the amplitude and strong phase variation over the Dalitz plot. 
This comes at the cost of a degree of ambiguity in the choice of test statistic quantifying the compatibility of the data with a given value of $\gamma$, with improved choices of test statistic yielding higher sensitivity.
While our current proof-of-principle implementation does not demonstrate optimal sensitivity to $\gamma$, its conceptually novel approach opens the door to new strategies for $\gamma$ extraction. 
More studies are required to see if these can be competitive with the existing methods.
\end{abstract}

\maketitle

\section{Introduction}
\label{sec:intro}

The $B$ meson decays $B^\pm \to D K^\pm$ are mediated through a combination of $b \to c\bar{u}s$ and $b \to u\bar{c}s$ transitions (and their conjugates).
Both receive tree-level contributions with comparable suppression factors from the Cabibbo--Kobayashi--Maskawa (CKM) matrix.
It has been recognized for over four decades that the interference of the resulting diagrams allows for efficient access to $CP$ violation in the CKM matrix~\cite{Carter:1980hr,Carter:1980tk}.
Specifically, these decay transitions provide sensitivity to the phase $\gamma$ (also referred to as $\phi_3$)~\cite{Gronau:1990ra,Gronau:1991dp} with negligible theoretical error~\cite{Brod:2013sga}, where
\begin{equation}
\label{eq:gdef}
    \gamma = \arg\pqty{-\frac{V_{ud} V_{ub}^*}{V_{cd} V_{cb}^*}}.
\end{equation}

The method of Refs.~\cite{Gronau:1990ra,Gronau:1991dp} uses two-body $D$ decays to $CP$ eigenstates, which is conceptually simple but suffers from suppressed sensitivity to $\gamma$ due to a large hierarchy between the two interfering amplitudes.
In Refs.~\cite{Atwood:1996ci,Atwood:2000ck}, this complication was ameliorated through the use of interference between Cabbibo-allowed and doubly-Cabbibo-suppressed flavor eigenstates of the neutral $D$ mesons.
However, the single best measurement of $\gamma$ is currently due to the BPGGSZ method~\cite{Bondar:2002, Giri:2003ty, Belle:2004bbr, Bondar:2005ki, Ceccucci:2020cim}, where $D^0$ and $\Dbar^0$ decay into multi-body $CP$ self-conjugate states (such as $K_S \pi^- \pi^+$) with large interference between the two parton-level weak transitions.
All of these methods have by now received many years of experimental effort~\cite{LHCb:2021dcr, LHCb:2020yot, BESIII:2020khq, Bjorn:2019kov, LHCb:2018wag, LHCb:2016aix, LHCb:2016bxi, LHCb:2014xrq, LHCb:2014kxv, BaBar:2013caj, LHCb:2012apu, Belle:2012ftx, BaBar:2010uep, Belle:2010xyn, CLEO:2009syb, BaBar:2008mds, BaBar:2008inr, Belle:2006lys, BaBar:2005rek, Bondar:2004bi, Belle:2004bbr}, with the resulting error on $\gamma$ at the several percent level.

In the BPGGSZ method, the $D$ decays are binned in the Dalitz plane in order to convert dependence on the full amplitude and phase into a finite number of bin-averaged values.
Symmetry properties of the $CP$-conjugate decay ensure that the number of independent quantities is smaller than the number of bins, provided the binning is symmetric with respect to $CP$ conjugation.
Given the leading role of the BPGGSZ method in the overall uncertainty of the measurement of $\gamma$, it is perhaps not surprising that the method has received significant experimental and theoretical attention, resulting in several improvements over the years.

Arguably the most significant is the optimized choice of binning, pioneered in Refs.~\cite{Bondar:2007ir,Bondar:2008hh}.
There, the shape of the bins in the $D \to K_S \pi^- \pi^+$ Dalitz plot is initialized to the isocontours of the strong phase difference $\Delta \delta_D = \delta_{13,12} - \delta_{12,13}$ (see \cref{eq:Damp,eq:Dbaramp} for definitions) between $CP$ symmetric points in the $D^0$ and $\bar D^0$ Dalitz plots. The bins are then continuously deformed in order to optimize sensitivity to $\gamma$.
This procedure minimizes the washout of sensitivity to $\gamma$ that occurs when $\Delta \delta_D$ is bin-averaged.
Binning in this manner is extremely powerful and is estimated to lead to only $\sim \SI{10}{\percent}$ lower statistical sensitivity to $\gamma$ for the choice of $2\times 8$ bins than if the unbinned amplitude model were used~\cite{Bondar:2008hh}.
(The amplitude model provides the smallest statistical error, but induces difficult-to-quantify systematical errors.)

An alternative approach, eliminating the bin dependence of the BPGGSZ method, was presented in Ref.~\cite{Poluektov:2017zxp}.
In that work, the binning of the $D$ decay Dalitz plot was replaced by a Fourier transform in phase space variables.
Truncation of the series gives a setup with a finite number of unknown parameters, such that it is possible to extract $\gamma$.
The choice of bins is thus replaced by the choice of variables in which to perform the Fourier transform, as well as of the order at which to truncate the series.
That is, some information from the Dalitz plane (in this case, the higher frequency components in the Fourier expansion) is still removed at the data processing step. 
Moreover, choosing a coordinate basis for the Fourier transform is driven by the same modeling of the strong phase that is used in the determination of optimized Dalitz plot bins.
While the Fourier transform method has not yet been implemented on experimental data, the study of Ref.~\cite{Poluektov:2017zxp} indicates that it can be highly competitive with the binned BPGGSZ method.

In this manuscript, we present a proof of principle for a novel $\gamma$ extraction method that requires neither binning nor truncation.
Given the highly optimized binning procedure currently used in experimental implementations of the BPGGSZ method, our main goal here is conceptual.
Our hope is that, eventually, this new method will be able to improve on the BPGGSZ method and/or the method of Ref.~\cite{Poluektov:2017zxp}.
This is because the existing methods all require removal of a certain amount of information during data processing.
While this step can be optimized, the lost information cannot be recovered by any statistical treatment.
We replace this with an alternative data processing procedure that, in principle, removes no decay information.
(Note, however, that not all relevant information is used in the current implementation.)
Our method makes the problem of $\gamma$ determination equivalent to the question of whether two inequivalent measurements are sampling the same function.
This structure allows for the application of a wide range of nonparametric methods to the problem, replacing the optimization of the data processing step with the conceptually different procedure of test statistic optimization.

The explicit analysis we introduce in the main part of the paper indeed falls short in terms of its statistical sensitivity to $\gamma$ compared to the methods above.
Nevertheless, we find that a toy example of a method without binning or truncation is still useful, as it makes clear that such approximations can be replaced with the problem of finding an optimal test statistic (that is, an observable optimally sensitive to $\gamma$). 
We wish to stress that while each of the available methods --- BPGGSZ, Ref.~\cite{Poluektov:2017zxp}, and the methods presented in this paper --- require some optimization in order to reduce the statistical error on $\gamma$, these optimizations take on qualitatively different forms.
For the BPGGSZ method, this is the number of bins and their shapes.
For Ref.~\cite{Poluektov:2017zxp}, it is the choice of Fourier transform variables and the order of truncation.
For our method, it is the choice of test statistic.
The question of which method can ultimately give the smallest statistical error on measurements of $\gamma$ depends critically on this optimization step.
We hope to improve on our proof-of-principle implementation with a follow-up work in which we perform such an optimization for our method.

The paper is structured as follows.
In \cref{sec:notation}, we review the dependence of the $B^\pm \to (K_S\pi^-\pi^+)_DK^\pm$ decay rates on the CKM angle $\gamma$. 
In \cref{subsec:partial}, we present a carefully chosen combination of reduced differential partial decay widths in order to extract $\cot^2\gamma$. 
In \cref{sec:strategy}, we derive the implications for cumulative reduced partial decay widths and present the fundamental idea of our algorithm.
Namely, we extract $\gamma$ as a parameter that brings two functions of empirical cumulative probability distributions into as much agreement as possible. 
We implement this idea by employing a measure adapted from the Kolmogorov--Smirnov test statistic in \cref{sec:KS-test}.
To demonstrate our proof of principle, we show numerical results based on toy Monte Carlo data in \cref{sec:results}. 
Conclusions are given in \cref{sec:conclusions}.

\section{Theory}
\label{sec:ginfer}

\subsection{Notation}
\label{sec:notation}

To set the stage, we review how sensitivity to the CKM unitarity triangle angle $\gamma$ is achieved in $B^\pm \to DK^\pm$ transitions, following the notation of Ref.~\cite{Giri:2003ty}.
Consider the $CP$-conjugate cascade decays
\begin{equation}
\label{eq:modes}
    B^\pm \to DK^\pm \to (K_S \pi^- \pi^+)_D K^\pm.
\end{equation}
A Dalitz plot analysis of $K_S \pi^- \pi^+$, characterizing the intermediate neutral $D$ meson, allows us to fully specify the $\gamma$-dependence of the process.
Focusing first on the initial two-body decay, we define
\begin{gather}
\label{eq:BDKnor}
  \A(B^- \to D^0 K^-) = \A(B^+ \to \Dbar^0 K^+) \equiv A_B, \\
\label{eq:BDKr}
  \A(B^- \to \Dbar^0 K^-) \equiv A_B r_B e^{i(\delta_B - \gamma)}, \qquad
  \A(B^+ \to D^0 K^+) \equiv A_B r_B e^{i(\delta_B + \gamma)}.
\end{gather}
$A_B$ is real by convention, with $\delta_B$ the difference between the strong phases of the $D^0 K$ and $\Dbar^0 K$ amplitudes.
The amplitudes in \cref{eq:BDKr} carry a weak phase which agrees with $\gamma$ in \cref{eq:gdef} up to $\ord(\lambda^4)$ corrections.
These have been computed to give relative shifts in the determination of $\gamma$ of $\sim \num{2e-3}$, which we ignore going forward~\cite{Brod:2013sga}. 
Due to the color and CKM suppression of the amplitudes in \cref{eq:BDKr}, the theoretical expectation is that $r_B$ is small ($r_B \sim \numrange{0.1}{0.2}$), in agreement with the experimental determination~\cite{LHCb:2021dcr},
\begin{equation}
\label{eq:rB-experimental}
  r_B = \num{0.0984}^{+0.0027}_{-0.0026}\,.
\end{equation}
The smallness of $r_B$ reduces sensitivity to $\gamma$ in all methods using $B^\pm \to DK^\pm$.
In the following analysis, we neglect $D^0$--$\Dbar^0$ mixing, which contributes at second order in the mixing parameters and yields a subleading correction of less than \SI{1}{\percent} in the extraction of $\gamma$~\cite{Grossman:2005rp}.

For the subsequent three-body decay of the intermediate neutral $D$ meson,
\begin{equation}
\label{eq:D-decay}
    D \to K_S(p_1) \pi^-(p_2) \pi^+(p_3),
\end{equation}
we define the amplitudes
\begin{align}
\label{eq:Damp}
    \A(D^0 \to K_S \pi^- \pi^+)
    &\equiv A(s_{12}, s_{13}) e^{i\delta(s_{12}, s_{13})} \equiv A_{12,13} e^{i\delta_{12,13}}, \\
\label{eq:Dbaramp}
    \A(\Dbar^0 \to K_S \pi^- \pi^+)
    &\equiv A(s_{13}, s_{12}) e^{i\delta(s_{13}, s_{12})} \equiv A_{13,12} e^{i\delta_{13,12}},
\end{align}
where $s_{ij} = (p_i + p_j)^2$ is the invariant mass squared of the $ij$ system.
We define $A_{12,13}$ and $\delta_{12,13}$ to be real functions with $A_{12,13} \geq 0$ and $\delta_{12,13}\in [0, 2\pi)$. 
The simple relationship between the amplitudes in \cref{eq:Damp,eq:Dbaramp} follows from the $CP$ symmetry of the strong interaction and the fact that the final state $K_S \pi^- \pi^+$ has zero spin. ($CP$ violation in $D$ decays is very small and thus is neglected in this discussion.)

The $D$ meson is a narrow state that decays weakly, justifying the use of the narrow width approximation and the neglect of any continuum contribution.
In the vicinity of the $D$ resonance, we thus have
\begin{align}
\label{eq:Bminus-amp}
  \A(B^- \to (K_S \pi^- \pi^+)_D K^-)
    &= A_B P_D \bqty{A_{12,13}e^{i\delta_{12,13}}
                     + A_{13,12} r_B e^{i(\delta_B - \gamma + \delta_{13,12})}}, \\
\label{eq:Bplus-amp}
  \A(B^+ \to (K_S \pi^- \pi^+)_D K^+)
    &= A_B P_D \bqty{A_{13,12}e^{i\delta_{13,12}}
                     + A_{12,13} r_B e^{i(\delta_B + \gamma + \delta_{12,13})}},
\end{align}
where $P_D$ is the neutral $D$ meson propagator.
In the narrow width approximation, we can write the $B$ meson partial width for these decays in terms of \cref{eq:Damp,eq:Dbaramp} as
\begin{equation}
  \frac{\dd{\Gamma_\pm}}{\dd{s_{12}}\dd{s_{13}}}
    = \frac{1}{(2\pi)^4} \frac{1}{128 m_D^3} \frac{\abs{\vec{p}_{K^\pm}}}{m_B^2 \Gamma_D}
        \abs{\A\bqty{D^0(\Dbar^0)} + r_B e^{i(\delta_B\mp\gamma)} \A\bqty{\Dbar^0(D^0)}}^2
\end{equation}
using
\begin{equation}
\label{eq:D-prop}
  P_D^2 = \frac{\pi}{m_D \Gamma_D} \delta(s_{123} - m_D^2),
\end{equation}
where $s_{123} = (p_1 + p_2 + p_3)^2$ denotes the invariant mass squared of the $K_S \pi^- \pi^+$ system.
Since the $\A(D)$-independent prefactors are the same for all decays, we can ignore them by defining reduced partial decay widths
\begin{align}
\label{eq:Bm-dal}
    \frac{\dd{\hat{\Gamma}_-}}{\dd{s_{12}} \dd{s_{13}}}
      &= A_{12,13}^2 + r_B^2 A_{13,12}^2
         + 2r_B A_{12,13}A_{13,12}\cos{(\delta_B - \gamma + \delta_{13,12} - \delta_{12,13})}, \\
\label{eq:Bp-dal}
    \frac{\dd{\hat{\Gamma}_+}}{\dd{s_{12}} \dd{s_{13}}}
      &= A_{13,12}^2 + r_B^2 A_{12,13}^2
         + 2r_B A_{12,13}A_{13,12}\cos{(\delta_B + \gamma +  \delta_{12,13} - \delta_{13,12})},
\end{align}
in agreement with Ref.~\cite{Giri:2003ty}.
Note that $\hat{\Gamma}$ is dimensionless.

We can obtain additional information about the magnitudes $A_{12,13}$ and strong phases $\delta_{12,13}$ from $D$ decay data.
In particular, $D^{*+} \to D^0\pi^+$ decays tell us about $A_{12,13}$.
Since data on $D \to K_S\pi^-\pi^+$ decays is abundant, this gives rather precise, direct information on $A_{12,13}$.
Measurements of the $D \to K_S\pi^-\pi^+$ Dalitz plot distributions in coherent $\psi(3770)\to D\Dbar$ decays, where at least one of the two $D$ decays is in the $K_S\pi^-\pi^+$ channel, give model-independent information about $\delta_{12,13}$ (though limited in statistics).
Still, at the moment, determinations of $D$ decay parameters are a subdominant source of error, almost an order of magnitude smaller than the statistical error incurred with the binned extraction of $\gamma$~\cite{LHCb:2020yot,BESIII:2020hlg,BESIII:2020khq}.

In our proof-of-principle unbinned method for $\gamma$ extraction (introduced in \cref{subsec:partial}), we will make two simplifications.
First, we will assume that the cumulative distributions of the $A_{12,13}$ functions are measured precisely enough to be treated as exactly known.
Secondly, we will formulate the method such that information about the strong phases $\delta_{12,13}$ is never required.
The first simplification is made for ease of presentation: in the intermediate theory expressions, we can treat $A_{12,13}$ as known functions, while in the final expressions only the cumulative distribution functions will be used (and errors on their determinations will certainly be subleading).
We would prefer to relax the second simplification in the future, since we are clearly discarding useful information.

\subsection{\boldmath \texorpdfstring{$\gamma$}{Gamma} from reduced partial widths in theory}
\label{subsec:partial}

We now demonstrate that, by considering simple odd and even combinations of the partial widths of \cref{sec:notation}, the relative dependence on the parameters of the $B^\pm \to DK^\pm$ decays take on a particularly simple form.
(For an alternative use of symmetry considerations to parameterize approximate flavor symmetries in 3-body decays, see Ref.~\cite{Bertholet:2018tmx}.)
With respect to the $s_{12} = s_{13}$ axis of the Dalitz plane, we first form even ($\Sigma$) and odd ($\Delta$) quantities made from the widths $\dd{\hat{\Gamma}_\pm}$ given in \cref{eq:Bm-dal,eq:Bp-dal}:
\begin{equation}
\label{eq:def-Sigma-Delta}
    \dd{\Sigma_\pm}(s_{12}, s_{13}),\ \dd{\Delta_\pm}(s_{12}, s_{13})
      \equiv \frac{\dd{\hat{\Gamma}_\pm}(s_{12}, s_{13}) \pm \dd{\hat{\Gamma}_\pm}(s_{13}, s_{12})}{2}.
\end{equation}
Here, the subscripts correspond to the $B^\pm$ charge and agree across both sides of the equation.
Further symmetrizing and anti-symmetrizing the quantities in \cref{eq:def-Sigma-Delta} with respect to the $B$ meson charge yields
\begin{align}
\label{eq:def-dD-dSigma}
  \dd{\Sigma_{S,A}}(s_{12}, s_{13})
    &\equiv \frac{\dd{\Sigma_+}(s_{12}, s_{13}) \pm \dd{\Sigma_-}(s_{12}, s_{13})}{2}, \\
\label{eq:def-dD-dDelta}
  \dd{\Delta_{S,A}}(s_{12}, s_{13})
    &\equiv \frac{\dd{\Delta_+}(s_{12}, s_{13}) \pm \dd{\Delta_-}(s_{12}, s_{13})}{2}.
\end{align}
The subscripts $S$ and $A$ on the left-hand sides of \cref{eq:def-dD-dSigma,eq:def-dD-dDelta} stand for \emph{symmetric} and \emph{anti-symmetric} and correspond to the choices of $+$ and $-$, respectively.
Inserting the explicit expressions given in \cref{eq:Bm-dal,eq:Bp-dal}, we find
\begin{align}
\label{eq:dSigmaS}
  \frac{\dd{\Sigma_S}}{\dd{s_{12}}\dd{s_{13}}}
    &= \frac{1 + r_B^2}{2}(A_{12,13}^2 + A_{13,12}^2)
       + 2r_B A_{12,13}A_{13,12}\cos{(\delta_{13,12} - \delta_{12,13})} \cos{\delta_B}\cos{\gamma}, \\
\label{eq:dSigmaA}
  \frac{\dd{\Sigma_A}}{\dd{s_{12}}\dd{s_{13}}}
    &= -2r_B A_{12,13}A_{13,12}\cos{(\delta_{13,12} - \delta_{12,13})} \sin{\delta_B} \sin{\gamma}, \\
\label{eq:dDeltaS}
  \frac{\dd{\Delta_S}}{\dd{s_{12}}\dd{s_{13}}}
    &= 2 r_B A_{12,13}A_{13,12} \sin{(\delta_{13,12} - \delta_{12,13})}\cos{\delta_B} \sin{\gamma}, \\
\label{eq:dDeltaA}
  \frac{\dd{\Delta_A}}{\dd{s_{12}}\dd{s_{13}}}
    &= \frac{1 - r_B^2}{2}(A_{13,12}^2 - A_{12,13}^2)
       + 2r_B A_{12,13} A_{13,12} \sin{(\delta_{13,12} - \delta_{12,13})} \sin{\delta_B} \cos{\gamma}.
\end{align} 
We finally define the quantities $\dd{\Sigma_S} |_\mathrm{sub}$ and $\dd{\Delta_A} |_\mathrm{sub}$ by subtracting away the first terms in \cref{eq:dSigmaS,eq:dDeltaA}, resulting in
\begin{align}
\label{eq:dSigmaS-sub}
\begin{split}
    \eval{\frac{\mathrm{d}\Sigma_S}{\dd{s_{12}}\dd{s_{13}}}}_\mathrm{sub} &\equiv \frac{\mathrm{d}\Sigma_S}{\dd{s_{12}}\dd{s_{13}}} - \frac{1 + r_B^2}{2}(A_{12,13}^2 + A_{13,12}^2)\\
    &= 2r_B A_{12,13}A_{13,12}\cos{(\delta_{13,12} - \delta_{12,13})} \cos{\delta_B}\cos{\gamma}, 
    \end{split}
    \\
\label{eq:dDeltaA-sub}
\begin{split}
    \eval{\frac{\mathrm{d}\Delta_A}{\dd{s_{12}}\dd{s_{13}}}}_\mathrm{sub} &\equiv \frac{\mathrm{d}\Delta_A}{\dd{s_{12}}\dd{s_{13}}} -\frac{1 - r_B^2}{2}(A_{13,12}^2 - A_{12,13}^2) \\
    &= 2r_B A_{12,13} A_{13,12} \sin{(\delta_{13,12} - \delta_{12,13})} \sin{\delta_B} \cos{\gamma}.
    \end{split}
\end{align}
Note that the subtracted quantities above are not directly observed experimentally.
However, while the terms subtracted away from \cref{eq:dSigmaS,eq:dDeltaA} depend on the $D$ decay amplitudes and $r_B$, they do not depend on $\gamma$.
Moreover, the subtracted terms only depend on $r_B$ quadratically, with the effect of subtracting using an incorrectly determined value of $r_B$ suppressed.
As explained above, for our proof-of-principle demonstration, we treat $A_{12,13}$ as a known function of $s_{12}$, $s_{13}$, determined with good enough precision from $D$ decay data, while $r_B$ is a parameter which we fit within the method.
Note further that the above subtraction is equivalent to replacing the decay widths given in \cref{eq:Bm-dal,eq:Bp-dal} by 
\begin{align}
\label{eq:gamma-sub-1}
  \eval*{\frac{\dd{\hat{\Gamma}_-}}{\dd{s_{12}} \dd{s_{13}}}}_\mathrm{sub}
    &\equiv \frac{\dd{\hat{\Gamma}_-}}{\dd{s_{12}} \dd{s_{13}}} - (A_{12,13}^2 + r_B^2 A_{13,12}^2),\\
\label{eq:gamma-sub-2}
  \eval*{\frac{\dd{\hat{\Gamma}_+}}{\dd{s_{12}} \dd{s_{13}}}}_\mathrm{sub}
    &\equiv \frac{\dd{\hat{\Gamma}_+}}{\dd{s_{12}} \dd{s_{13}}} - (A_{13,12}^2 + r_B^2 A_{12,13}^2), 
\end{align}
and then forming combinations of them analogous to  \cref{eq:def-Sigma-Delta,eq:def-dD-dSigma,eq:def-dD-dDelta}.

It is easy to see that the ratios
\begin{align}
\label{eq:sub-main-1}
   \eval{\frac{\mathrm{d}\Sigma_S}{\dd{s_{12}}\dd{s_{13}}}}_\mathrm{sub}
     \Bigg/ \frac{\mathrm{d}\Sigma_A}{\dd{s_{12}}\dd{s_{13}}}
       &= -\cot{\delta_B}\cot{\gamma}\\
\label{eq:sub-main-2}
   \eval{\frac{\mathrm{d}\Delta_A}{\dd{s_{12}}\dd{s_{13}}}}_\mathrm{sub}
     \Bigg/ \frac{\mathrm{d}\Delta_S}{\dd{s_{12}}\dd{s_{13}}}
       &= \tan{\delta_B} \cot{\gamma},
\end{align}
take on constant values in the Dalitz plane. 
Furthermore, the product of these two ratios,
\begin{equation}
\label{eq:sub-prod}
  \pqty{\eval{\frac{\mathrm{d}\Sigma_S}{\dd{s_{12}}\dd{s_{13}}}}_\mathrm{sub}
        \eval{\frac{\mathrm{d}\Delta_A}{\dd{s_{12}}\dd{s_{13}}}}_\mathrm{sub}}
    \Bigg/ \pqty{\frac{\mathrm{d}\Sigma_A}{\dd{s_{12}}\dd{s_{13}}}  
                 \frac{\mathrm{d}\Delta_S}{\dd{s_{12}}\dd{s_{13}}}}
      = -\cot^2{\gamma},
\end{equation}
allows us direct access to $\gamma$  up to a four-way degeneracy equivalent to that in Ref.~\cite{Giri:2003ty}.
The relations in \cref{eq:sub-main-1,eq:sub-main-2} are what form the basis of the unbinned method for extracting $\gamma$ described in the subsequent sections of this work.

\subsection{\boldmath \texorpdfstring{$\gamma$}{Gamma} from cumulative reduced partial widths in practice}
\label{sec:strategy}

Our observations consist of individual $B$ decay events and not continuous distributions.
To make connections with the results above, we introduce the cumulative reduced partial decay widths, defined as\footnote{This is just one possible definition of the cumulative reduced partial decay widths. A priori, other integration orders are equally viable.}
\begin{equation}
\label{eq:cdfdef}
  R_\pm(s_{12}, s_{13}) \equiv \int_0^{s_{12}} \dd{s_{12}'} \int_0^{s_{13}} \dd{s_{13}'}
                                 \frac{\dd{\hat{\Gamma}_\pm}}{\dd{s_{12}'}\dd{s_{13}'}},
\end{equation}
where, outside the physical region of the Dalitz plot, $\dd{\hat{\Gamma}_\pm}/(\dd{s_{12}} \dd{s_{13}})= 0$.
These functions are monotonically increasing, with limiting values $R_\pm(0, 0) = 0$ and $R_\pm(s_{12}, s_{13} > (m_D - m_\pi)^2) = \hat{\Gamma}_\pm^\text{(tot)}$, where
\begin{equation}
  \hat{\Gamma}_\pm^\text{(tot)} = \int \dd{\hat{\Gamma}_\pm} (s_{12}, s_{13})\,
\end{equation}
is the reduced partial width of $B^\pm \to DK^\pm$.
Note that $\hat{\Gamma}_\pm^\text{(tot)}$ is not, in general, normalized to unity, and, thus, resulting objects do not directly correspond to cumulative distribution functions.
The functions $R_\pm(s_{12}, s_{13})$ continue to vary outside of the physical Dalitz region, \ie, in the rectangle around the physical region of the Dalitz plot.

With many observed $B^\pm$ decay events, \cref{eq:cdfdef} will be approached, up to an overall constant, by the counting functions
\begin{equation}
\label{eq:cumcount}
    N_\pm(s_{12}, s_{13}) = \sum_{\substack{i_\pm < s_{12}\\j_\pm < s_{13}}} 1\,,
\end{equation}
where the index $i_{\pm}$ ($j_{\pm}$) runs over the $s_{12}$ ($s_{13}$) values of all observed decays.
$N_\pm(s_{12}, s_{13})$ are thus the number of observed $B^\pm \to DK^\pm$ decays with $(p_K + p_{\pi^-})^2 < s_{12}$ and $(p_K + p_{\pi^+})^2 < s_{13}$.
They are nonzero almost everywhere in the Dalitz plane and can be constructed with no binning and with minimal processing of the data.
These functions are approximations of the continuous partial decay widths, such that, in the limit of large $N_\pm^{\text{(tot)}}$, the total number of $B^\pm \to DK^\pm$ events, we have
\begin{equation}
  \frac{R_{\pm}(s_{12}, s_{13})}{\hat{\Gamma}_\pm^\text{(tot)}}
    \approx \frac{N_{\pm}(s_{12}, s_{13})}{N_\pm^\text{(tot)}}\,.
\end{equation}

We now bring \cref{eq:cumcount} through the procedure laid out in \cref{subsec:partial}, namely, we symmetrize and anti-symmetrize the function with respect to phase space and initial $B$ meson charge.
This gives
\begin{align}
\label{eq:R-S-D}
    R_{\Sigma S, \Sigma A}(s_{12}, s_{13}) &= \frac{1}{2}(\Sigma_+(s_{12}, s_{13}) \pm \Sigma_-(s_{12}, s_{13})), \\
\label{eq:R-D-D}
    R_{\Delta S, \Delta A}(s_{12}, s_{13}) &= \frac{1}{2}(\Delta_+(s_{12}, s_{13}) \pm \Delta_-(s_{12}, s_{13})),
\end{align}
with
\begin{equation}
\label{eq:S-D}
    \Sigma_\pm(s_{12}, s_{13}),\  \Delta_\pm(s_{12}, s_{13})
      = \frac{N_{\pm}(s_{12}, s_{13}) \pm N_{\pm}(s_{13}, s_{12})}{2},
\end{equation}
which are the cumulative versions of the functions in \cref{eq:def-Sigma-Delta,eq:def-dD-dSigma,eq:def-dD-dDelta}. 
We may also find the cumulative versions of \cref{eq:dSigmaS-sub,eq:dDeltaA-sub} by subtracting away the appropriate cumulative flavor-tagged $D^0$ meson decays. 
To do this, we define 
\begin{align}
\label{eq:Nminus-imp}
    N_-(s_{12}, s_{13})|_\mathrm{sub}
      &= N_-(s_{12}, s_{13}) - r_-\frac{N_-^\text{(tot)}}{N_D^\text{(tot)}}
                                 (N_D(s_{12}, s_{13}) + r_B^2 N_D(s_{13}, s_{12})),\\
\label{eq:Nplus-imp}
    N_+(s_{12}, s_{13})|_\mathrm{sub}
      &= N_+(s_{12}, s_{13}) - r_+\frac{N_+^\text{(tot)}}{N_D^\text{(tot)}}
                                 (N_D(s_{13}, s_{12}) + r_B^2 N_D(s_{12}, s_{13})), 
\end{align}
where
\begin{equation}
\label{eq:rpm-rat}
    r_\pm = \frac{\int \dd{s_{12}}\dd{s_{13}} |\A(D^0 \to K_S \pi^- \pi^+)|^2}
                 {\hat{\Gamma}_\pm^\text{(tot)}},
\end{equation}
is necessary for the proper normalization of \cref{eq:Nplus-imp,eq:Nminus-imp}.
The forms of the subtracted parts in \cref{eq:Nminus-imp,eq:Nplus-imp} follow from \cref{eq:gamma-sub-1,eq:gamma-sub-2}.
Here, $N_D^\text{(tot)}$ is the total number of (independently) observed  $D^0 \to K_S \pi^- \pi^+$ events, while $N_D(s_{12}, s_{13})$ is defined in exactly the same way as $N_\pm(s_{12}, s_{13})$ in \cref{eq:cumcount} except with summations now over all $D^0 \to K_S \pi^- \pi^+$ events.

Note that the normalization factor $r_{\pm}$ in \cref{eq:rpm-rat} is measured in the experiment and is simply given by the ratio of measured $D^0 \to K_S\pi^-\pi^+$ and (reduced) $B^\pm \to (K_S\pi^-\pi^+)_D K^\pm$ decay rates.
For the purposes of our proof-of-principle demonstration, we set $r_\pm$ to its expected measured value as predicted by $\gamma$, $\delta_B$, and $r_B$ values and do not take into account experimental errors.
Likewise, for the purposes of the calculation of $r_\pm$, we use the $D^0 \to K_S\pi^-\pi^+$ amplitude model of Ref.~\cite{BaBar:2018cka}, including the strong phases.
However, in the algorithm itself, we use the discrete data $N_D(s_{12},s_{13})$ only and no phase information.
The experimental determination of $N_D(s_{12},s_{13})$ can make use of both $D^0$ and $\Dbar^0$ decays, since, in the above discussion, we neglect $CP$ violation in $D$ decays. Note also that the discussion of experimental effects --- such as backgrounds, efficiencies, and resolutions --- is outside of the scope of the present manuscript. 

In the next step, we use \cref{eq:Nminus-imp,eq:Nplus-imp} in \cref{eq:R-S-D,eq:R-D-D,eq:S-D} to get the cumulative functions $R_{\Sigma S}|_\mathrm{sub}$ and $R_{\Delta A}|_\mathrm{sub}$.
Due to the linearity of the integral operation in \cref{eq:cdfdef}, the relations in \cref{eq:sub-main-1,eq:sub-main-2,eq:sub-prod} all hold if we replace the functions therein with their corresponding cumulative counterparts.
That is, we have
\begin{equation}
\label{eq:sub-R}
    \frac{R_{\Sigma S}|_\mathrm{sub}}{R_{\Sigma A}} = -\cot\delta_B \cot\gamma \qc
    \frac{R_{\Delta A}|_\mathrm{sub}}{R_{\Delta S}} = \tan\delta_B \cot\gamma,
\end{equation} 
and
\begin{equation}
\label{eq:prod-R}
    \left(\frac{R_{\Sigma S}|_\mathrm{sub}}{R_{\Sigma A}}\right)
    \left(\frac{R_{\Delta A}|_\mathrm{sub}}{R_{\Delta S}}\right) = -\cot^2 \gamma.
\end{equation}
Stopping to examine the ratios in \cref{eq:sub-R} closely, we emphasize the following observation:
Up to a $\delta_B$- and $\gamma$-dependent rescaling, the cumulative functions within each $R_{\Sigma}$ and $R_{\Delta}$ pair are the same.
For instance, note that the first equation in \cref{eq:sub-R} tells us that $R_{\Sigma S}|_{\rm{sub}} = -\cot\delta_B \cot\gamma\, R_{\Sigma A}$, where the left-hand side depends only on a correct determination of $r_B$ (due to the subtraction, see the first line in~\cref{eq:dSigmaS-sub}) and the right-hand side only on $\delta_B$ and $\gamma$.
As a result, we recast the problem of measuring $\gamma$ as one of finding values of $\gamma$, $\delta_B$, and $r_B$ that give the two pairs of functions $R_\Sigma$ and $R_\Delta$, rescaled following \cref{eq:sub-R}, the highest statistical significance of having been drawn from the same underlying distributions.
Note that, below, we make use of \cref{eq:sub-R} rather than \cref{eq:prod-R}, since \cref{eq:sub-R} allows for the extraction of all three parameters $\gamma$, $\delta_B$, and $r_B$, while \cref{eq:prod-R} is sensitive only to $\gamma$ and $r_B$.

\section{\boldmath Extraction of \texorpdfstring{$\gamma$}{Gamma} as an Optimization Problem}
\label{sec:KS-test}

The strategy of \cref{sec:strategy} for extracting $\gamma$, $\delta_B$, and $r_B$ is condensed in \cref{eq:sub-R}.
One way to practically employ this equation is to vary these three parameters such that the functions
\begin{align}
\label{eq:ansatz-1}
  D_\Sigma(\gamma, \delta_B, r_B) &\equiv
    \max_{s_{12},s_{13}} \big|R_{\Sigma S}|_\mathrm{sub}(s_{12},s_{13})
                               - (-\cot\delta_B \cot\gamma) R_{\Sigma A}(s_{12},s_{13})\big|, \\
\label{eq:ansatz-2}
  D_\Delta(\gamma, \delta_B, r_B) &\equiv
     \max_{s_{12},s_{13}} \big|R_{\Delta A}|_\mathrm{sub}(s_{12},s_{13})
                               - (\tan\delta_B \cot\gamma) R_{\Delta S}(s_{12},s_{13}) \big|,
\end{align}
are minimized.
The respective minima of $D_\Sigma(\gamma, \delta_B, r_B)$ and $D_\Delta(\gamma, \delta_B, r_B)$ should be reached for the same values of $\gamma$, $\delta_B$, and $r_B$, within statistical uncertainties.

This procedure is analogous to the minimization of the Kolmogorov-Smirnov (KS) test statistic.
In its original, one-dimensional formulation, the KS test takes two empirical cumulative distribution functions (CDFs), $F_{1}(x)$ and $F_{2}(x)$, and computes, as its test statistic, their maximum difference:
\begin{equation}
\label{eq:oneD-KS}
    D^\mathrm{KS} \equiv \max_x |F_{1}(x) - F_{2}(x)|.
\end{equation}
Two-dimensional realizations of the KS test, which are most relevant to our functions in \cref{eq:ansatz-1,eq:ansatz-2}, have been described in Refs.~\cite{Peacock:1983pi,Fasano:1987,Press:1988hk}.
Note importantly that \cref{eq:ansatz-1,eq:ansatz-2} are not exactly KS test statistics, as, unlike in the KS test, the functions $R_{\Sigma S}|_\mathrm{sub}$, $R_{\Sigma A}$, $R_{\Delta A}|_\mathrm{sub}$, and $R_{\Delta S}$ are not CDFs because they are not positive definite.

One complication that arises for two-dimensional cumulative functions is how to deal with the orientation of the integration in \cref{eq:cdfdef}. 
That is, including \cref{eq:cdfdef}, we may define $R_\pm(s_{12}, s_{13})$ in an infinite number of ways, each one an equally valid alternative to constructing test statistics $D_\Sigma^R(\gamma, \delta_B, r_B)$ and $D_\Delta^R(\gamma, \delta_B, r_B)$. 
This freedom in unbinned methods of extracting $\gamma$ corresponds to the infinite number of possible binnings of phase space in the binned methods.
In this present work, we limit our analysis to two additional definitions of cumulative $R_\pm(s_{12}, s_{13})$ functions, given by
\begin{align}
\label{eq:def-Rb}
  \widebar{R}_\pm(s_{12}, s_{13}) &\equiv
    \int_{s_{12}}^\infty\dd{s_{12}'} \int_0^{s_{13}} \dd{s_{13}'}
      \frac{\dd{\hat{\Gamma}_\pm}}{\dd{s_{12}'}\dd{s_{13}'}}, \\
\label{eq:def-Rt}
  \widetilde{R}_\pm(s_{12}, s_{13}) &\equiv
    \int_{s_{12}}^\infty \dd{s_{12}'} \int_{s_{13}}^\infty \dd{s_{13}'}
      \frac{\dd{\hat{\Gamma}_\pm}}{\dd{s_{12}'}\dd{s_{13}'}}.
\end{align}
We recall that, outside the Dalitz plot, $\dd{\hat{\Gamma}_\pm}/(\dd{s_{12}}\dd{s_{13}}) = 0$.
Generalizing \cref{eq:ansatz-1,eq:ansatz-2} in this fashion, this means that we additionally minimize
\begin{align}
\label{eq:DSigmaKS}
    \widetilde{D}_\Sigma(\gamma, \delta_B, r_B)
      &= \max_{s_{12}, s_{13}}  \abs{\widetilde{R}_{\Sigma S}|_\mathrm{sub}(s_{12}, s_{13})
           - (-\cot{\delta_B}\cot{\gamma}) \widetilde{R}_{\Sigma A}(s_{12}, s_{13})},  \\
\label{eq:DDeltaKS}
    \widetilde{D}_\Delta(\gamma, \delta_B, r_B)
      &= \max_{s_{12}, s_{13}}  \abs{\widetilde{R}_{\Delta A}|_\mathrm{sub}(s_{12}, s_{13})
          - (\tan{\delta_B}\cot{\gamma}) \widetilde{R}_{\Delta S}(s_{12}, s_{13}) }\,, 
\end{align}
as well as the analogously-defined $\widebar{D}_{\Sigma}(\gamma, \delta_B, r_B)$ and $\widebar{D}_{\Delta}(\gamma, \delta_B, r_B)$ test statistics.
In total, we therefore consider three out of the infinite different integration orderings, \ie, the orderings specified in \cref{eq:cdfdef,eq:def-Rb,eq:def-Rt}.
For finite data, the unbinned extraction of $\gamma$ works most effectively if one takes many more orderings into account and chooses the one that best minimizes the corresponding $D_\Sigma(\gamma, \delta_B, r_B)$ and $D_\Delta(\gamma, \delta_B, r_B)$ functions.

To compute the $\widebar{R}$ and $\widetilde{R}$ versions of $R_{\Sigma S}|_\mathrm{sub}$, $R_{\Sigma A}$, $R_{\Delta A}|_\mathrm{sub}$, and $R_{\Delta S}$, we follow the steps laid out in \cref{sec:strategy} but use, respectively, the following modified versions of the counting function in \cref{eq:cumcount}:
\begin{equation}
\label{eq:bt-cumcount}
    \widebar{N}_\pm(s_{12}, s_{13}) = \sum_{\substack{i_\pm > s_{12}\\j_\pm < s_{13}}} 1 \qc
    \widetilde{N}_\pm(s_{12}, s_{13}) = \sum_{\substack{i_\pm > s_{12}\\j_\pm > s_{13}}} 1.
\end{equation}
The forms of these functions follow trivially from the integrations in \cref{eq:def-Rb,eq:def-Rt}.
Using the $D$ meson decay data, we may also define $\widebar{N}_{D}(s_{12}, s_{13})$ and $\widetilde{N}_{D}(s_{12}, s_{13})$ in the same way as above for the computation of the $\widebar{R}$ and $\widetilde{R}$ versions of \cref{eq:Nminus-imp,eq:Nplus-imp}.

In summary, we extract a measurement of $\gamma$ as follows. Varying $\gamma$, $\delta_B$, and $r_B$, we determine the locations of the minima
\begin{align}
\label{eq:DminR}
  D_\mathrm{min} &=
    \min_{\gamma, \delta_B, r_B}\pqty{D_\Sigma^R(\gamma, \delta_B, r_B)
                                      + D_\Delta^R(\gamma, \delta_B, r_B)},\\
\label{eq:DminRtilde}
  \widetilde{D}_{\mathrm{min}} &=
    \min_{\gamma, \delta_B, r_B}\pqty{D_\Sigma^{\widetilde{R}}(\gamma, \delta_B, r_B)
                                      + D_\Delta^{\widetilde{R}}(\gamma, \delta_B, r_B)},\\
\label{eq:DminRbar}
  \widebar{D}_{\mathrm{min}} &=
    \min_{\gamma, \delta_B, r_B}\pqty{D_\Sigma^{\widebar{R}}(\gamma, \delta_B, r_B)
                                      + D_\Delta^{\widebar{R}}(\gamma, \delta_B, r_B)}.
\end{align}
As the minima of $D_\Sigma(\gamma, \delta_B, r_B)$ and $D_\Delta(\gamma, \delta_B, r_B)$ are at the same point for each choice of $R$, we combine the functions in the above fashion for symmetry reasons.
With infinite data, we would expect to achieve 
\begin{equation}
\label{eq:infinite-data}
  D_\mathrm{min} = \widetilde{D}_\mathrm{min} = \widebar{D}_\mathrm{min} = 0, 
\end{equation}
for a particular set of parameter values $\gamma$, $\delta_B$, and $r_B$.
With finite data, some integration orderings will achieve more effective minimizations in \cref{eq:DminR,eq:DminRtilde,eq:DminRbar}, meaning that deeper minima are attained.
We therefore identify the ``best'' of the integration orderings by the one that gives the deepest minimum, corresponding to the most reliable value of $\gamma$, which we quote as our final result.

\section{Numerical Results}
\label{sec:results}

\begin{figure}[t]
  \begin{center}
    \subfigure[]{\includegraphics[width=0.48\textwidth]{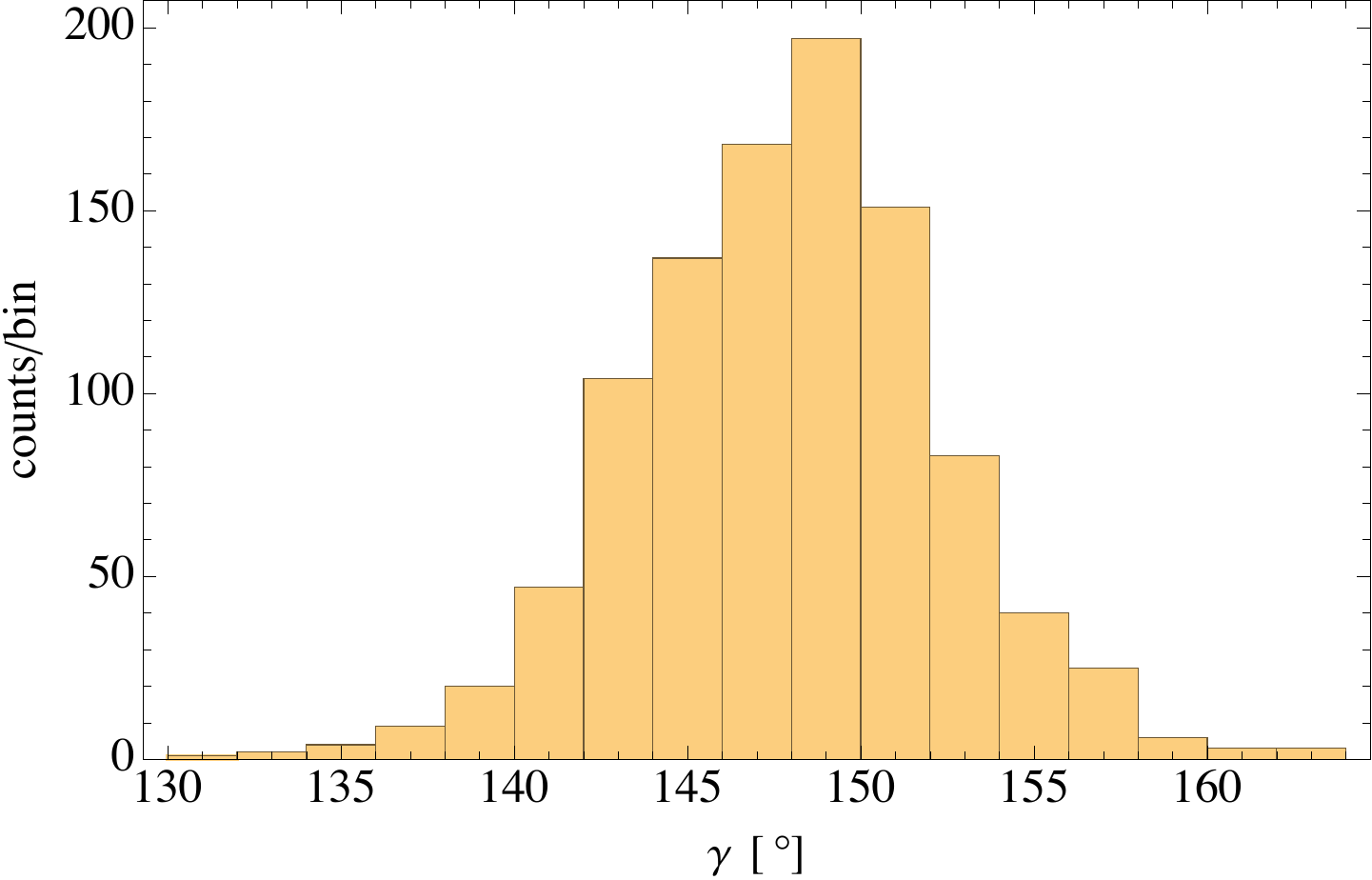}}
    \subfigure[]{\includegraphics[width=0.48\textwidth]{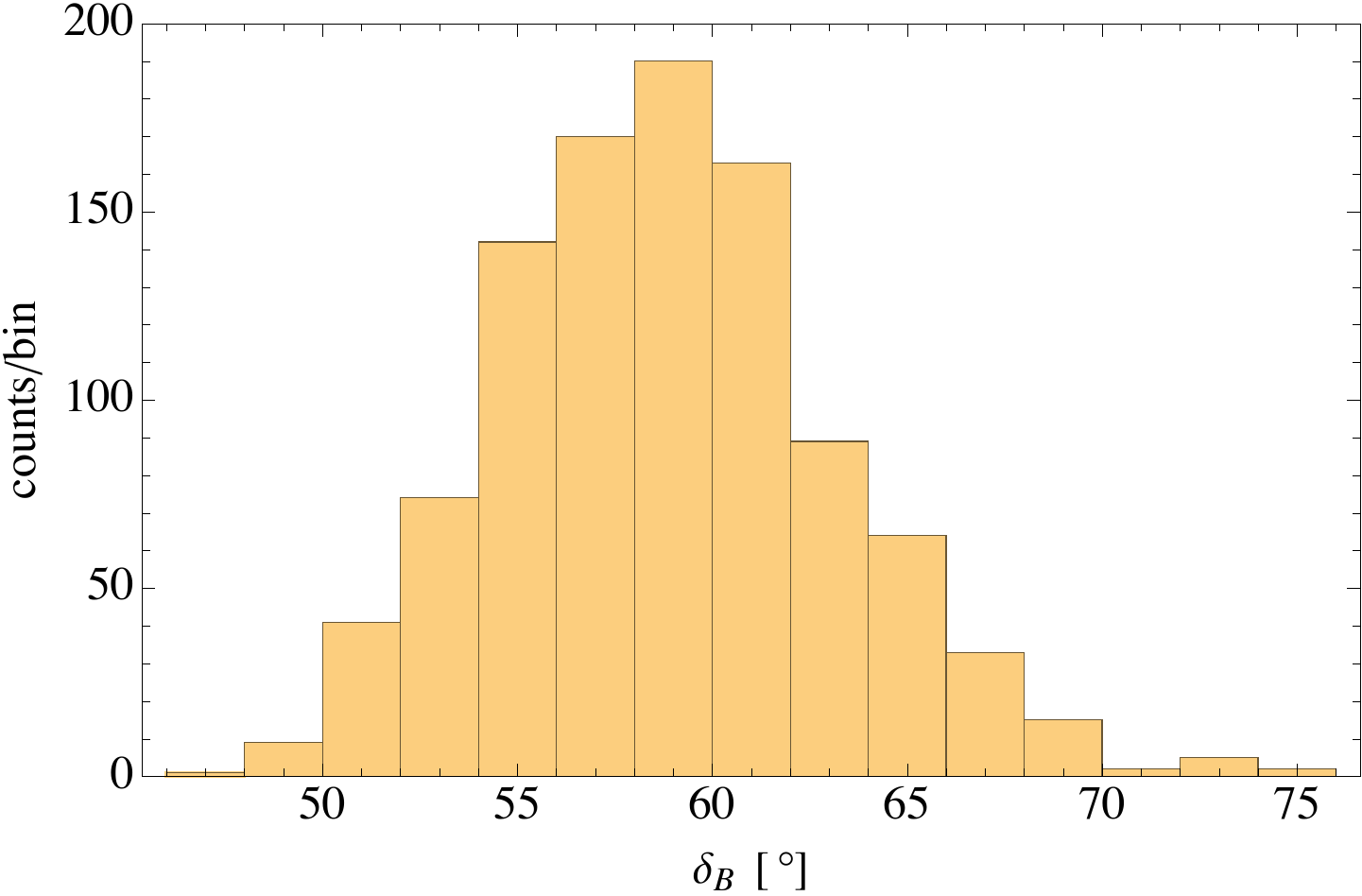}}
    \subfigure[]{\includegraphics[width=0.48\textwidth]{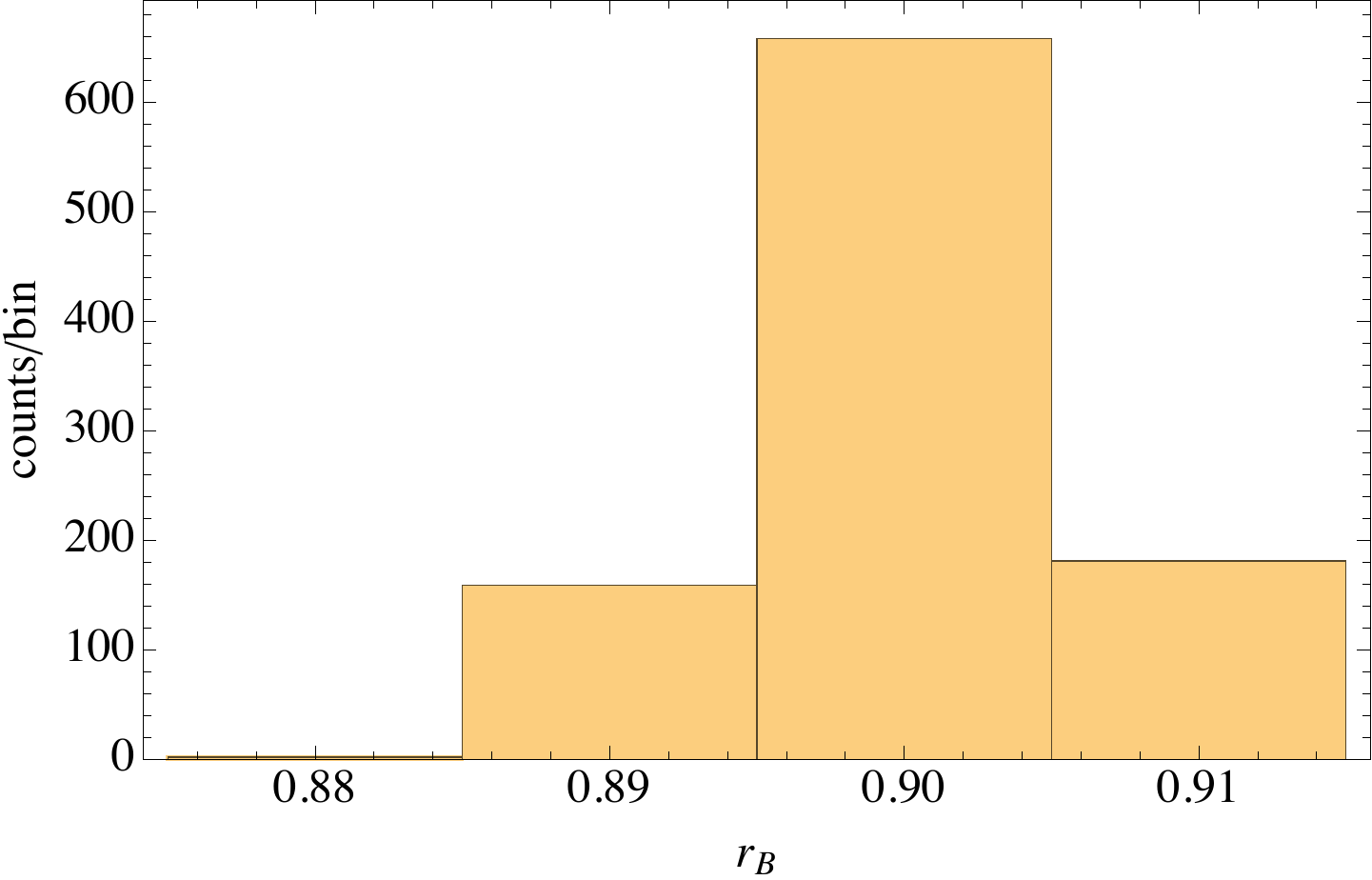}}
  \end{center}
  \caption{\label{fig:histograms-I}
     Histograms of the global minima for $\gamma$ (a), $\delta_B$ (b), and $r_B$ (c) obtained by applying our method to 1000 sets of toy Monte Carlo data. The input parameters are fixed to the values shown in \cref{tab:results}. Each of the Monte Carlo samples has \num{7000} $B^+$ decay events, \num{7000} $B^-$ decay events, and \num{5500} (independently-measured) $D$ decay events.}
\end{figure}

As a proof of principle of our unbinned methodology, we generate 1000 sets of toy Monte Carlo Dalitz plots with fixed input values for $\gamma$, $\delta_B$, and $r_B$ given in Table \ref{tab:results}. 
To do this, we implement the Dalitz plot amplitude model for $D \to K_S\pi^-\pi^+$ from Ref.~\cite{BaBar:2018cka}; see Refs.~\cite{Back:2017zqt, CLEO:2000fvk} for further details.\footnote{The Zemach factors~\cite{Zemach:1963bc, Zemach:1965ycj}, using the conventions of Ref.~\cite{CLEO:2000fvk}, have different mass dimension depending on the spin of the resonance. We assume therefore that the amplitude coefficients given in Table~III of Ref.~\cite{BaBar:2018cka} compensate the mass dimension of the Zemach factors. Note, however, that our proof of principle demonstration does not rely on the details of the amplitude model of the charm decay.}
We apply our unbinned procedure to each of these sets of generated data, arriving at 1000 extractions of the three parameters $\gamma$, $\delta_B$, and $r_B$ using \cref{eq:DminR,eq:DminRtilde,eq:DminRbar}.  Each of the Monte Carlo samples has \num{7000} $B^+$ decay events, \num{7000} $B^-$ decay events, and \num{5500} (independently-measured) $D$ decay events. Note that we use input values for $\gamma$, $\delta_B$, and $r_B$ that are quite far away from the values realized in nature. This is because the main goal of the present Monte Carlo study is to demonstrate that an unbinned extraction of $\gamma$ is possible. From the outset, it is clear that, in its present form, this model-independent unbinned method is much less sensitive to $\gamma$ than the optimized binned one is.

In order to identify the most effective integration ordering, we calculate the average values of $D_\mathrm{min}$, $\widetilde{D}_\mathrm{min}$, and $\widebar{D}_\mathrm{min}$ over these \num{1000} Dalitz plots and choose the smallest one. 
For the optimal integration ordering, we histogram the obtained global minima and extract our measurements of $\gamma$, $\delta_B$, and $r_B$ as the averages of these histograms.
The respective errors are given by the left and right bounds on the middle $\SI{68}{\percent}$ of entries in each histogram, resulting, in general, in asymmetric errors.
In future experimental analyses, this statistical treatment can be replaced by a more sophisticated procedure.

In order to find the global minimum of each set of generated data according to \cref{eq:DminR,eq:DminRtilde,eq:DminRbar}, we vary $\gamma$, $\delta_B$, and $r_B$ simultaneously.
We vary $\gamma$ in steps of $2^{\circ}$ in the interval~$[91^{\circ}, 179^{\circ}]$   and $\delta_B$ in steps of $2^{\circ}$ in the interval~$[1^{\circ}, 89^{\circ}]$.
That is, we search for the global minimum in the complete quadrant of the input values given in \cref{tab:results}.
Further, we vary $r_B$ in steps of $0.01$ in the interval~$[0.8, 1.0]$, corresponding to the value of $r_B = 0.9$ in \cref{tab:results}.
These choices were made because of the high computational cost of computing the functions in \cref{eq:ansatz-1,eq:ansatz-2} for $R$, $\widebar{R}$, and $\widetilde{R}$ (we perform simulations on a personal laptop), as well as some degeneracies that appear because \cref{eq:sub-R} only constrains trigonometric functions of $\gamma$ and $\delta_B$. For example, one such degeneracy occurs due to the fact that
\begin{equation}
\label{eq:degeneracy}
    \cot(\pi - \theta) = -\cot\theta \qc
    \tan(\pi - \theta) = -\tan\theta \,.
\end{equation}
Additionally, because it is impractical to implement a computation of \cref{eq:cumcount,eq:bt-cumcount} at all points $(s_{12}, s_{13})$, we instead sample the functions at a discrete set of points in the rectangle surrounding the physical region of the Dalitz plot and interpolate. 
In our implementation, we use grid points with a horizontal and vertical separation of \SI{0.01}{\GeV^2}.

\begin{table}[t]
  \centering
  \begin{tabular}{ccccc}
  \hline\hline
    ~Variable~ & ~Input~ & ~Output~ \\
    \hline
    $\gamma$ & $\num{150}^{\circ}$ &  $(148^{+3}_{-5})^{\circ}$ \\
    $r_B$ & \num{0.9} & $0.90^{+0.01}_{-0.01}$ ~ \\ 
    $\delta_B$ & $\num{60}^{\circ}$ & $(59^{+4}_{-4})^{\circ}$ \\
    \hline\hline
\end{tabular}
\caption{Input values for the parameters $\gamma$, $\delta_B$, and $r_B$ used for the generation of the toy Monte Carlo data and the corresponding output of the implementation of our unbinned algorithm.  \label{tab:results}}
\end{table}

We show the results of these computations for our considered scenario in \cref{tab:results}.
In this case, the optimal ordering turns out to be $R$.
In particular, although the average value of $\widetilde{D}_\mathrm{min}$ is nearly the same as that of $D_\mathrm{min}$, the average value of $\widebar{D}_\mathrm{min}$ is larger than that of $\widetilde{D}_\mathrm{min}$ and $D_\mathrm{min}$ by about an order of magnitude.
We give the resulting histograms of the 1000 extracted values of $\gamma$, $\delta_B$, and $r_B$ in \cref{fig:histograms-I}.
From \cref{tab:results}, we see that the output achieved by our unbinned analysis technique agrees with the input.

\section{Conclusions}
\label{sec:conclusions}

In this work, we have introduced a new, model-independent, unbinned method designed to extract $\gamma$ from $B^\pm \to DK^\pm \to (K_S \pi^- \pi^+)_D K^\pm$ decays.
This development contributes to the long-term effort towards achieving ultimate precision in the determination of $\gamma$, enabling unprecedented tests of the Standard Model. It is presently unclear if our method can provide a superior statistical error as compared to the other two theoretically clean methods. As mentioned in the introduction, each approach requires a different kind of statistical optimization. For us, the required optimization is in the choice of test statistic, which can be formed from variants of cumulative distribution functions such as Eq.~\eqref{eq:cdfdef}.

On the other hand, our method does not involve additional optimization of auxiliary variables that specify the analysis (such as shapes of bins or Fourier modes), unlike both the classic BPGGSZ method and the method of Ref.~\cite{Poluektov:2017zxp}. 
%
However, we are still in the early stages of developing a competitive alternative. 
Using toy Monte Carlo data, we have demonstrated as a proof of principle that this method returns values for $\gamma$, $\delta_B$, and $r_B$ that are in agreement with the input values chosen for the generated data. Future work is required to see how one can optimize the test statistic for this particular method.

Our method is not yet optimized for the highest sensitivity to $\gamma$ since it does not include all the possibly relevant observables.
For instance, by forming ratios such as those in \cref{eq:sub-main-1,eq:sub-main-1}, one reduces the number of observables sensitive to $\gamma$.
That not all of the available information is being used is further signaled by the fact that this method requires no information about the phases of the $D$ decay amplitudes, while these phases are an integral part of the more sensitive binned methods.
The competitiveness of the unbinned method could thus be greatly enhanced in the future by extending this approach to include data from correlated charm decays.

Just as the effectiveness of binned methods depends on the choice of the binning, the effectiveness of our unbinned method depends on the integration ordering of the cumulative functions.
In principle, there are an infinite number of ways to perform this integration over the Dalitz plot.
As such, it is not presently clear whether the binned or the unbinned methods will ultimately give the most competitive results.

\section*{Acknowledgments}
We thank Jared A.~Evans, Evelina Gersabeck, Tim Gershon, Jake Lane, Vincent Tisserand, Fernando Martinez-Vidal, and Julia Thom-Levy for useful discussions.
J.B. was supported in part by a Rawlings Cornell Presidential Research Scholarship.
M.F. was supported by the DOE under grant DE-SC0010008 and is supported by the NSF under grant PHY1316222. 
Y.G. is supported in part by the NSF grant PHY1316222.
S.S. is supported by a Stephen Hawking Fellowship from UKRI under reference EP/T01623X/1 and the Lancaster-Manchester-Sheffield Consortium for Fundamental Physics under STFC research grant ST/T001038/1.
J.Z. acknowledges support in part by the DOE grant DE-SC0011784 and NSF OAC-2103889.
For the purpose of open access, the authors have applied a Creative Commons Attribution (CC BY) licence to any Authors Accepted Manuscript version arising.
This work uses existing data which is available at locations cited in the bibliography. 

\bibliography{Gamma_biblio}

\end{document}